\documentclass[showpacs,preprint,aps]{revtex4}
\usepackage{graphicx}

\begin{document}
\setcounter{page}{1}
\title
{Electron tunneling times
\let\thefootnote\relax\footnote{Talk given at the 2$^{nd}$ Jagiellonian 
Symposium on Fundamental and Applied Subatomic Physics, Krak\'ow, 
June 4 - 9, 2017 }
}
\author 
{N. G. Kelkar}
\affiliation{ Departamento de Fisica, Universidad de los Andes, Cra 1E, 18A-10, 
Bogot\'a, Colombia}
\begin{abstract}
Tunneling is one of the most bizarre phenomena in quantum mechanics. An
attempt to understand it led to the next natural question of how long does
a particle need to tunnel a barrier. The latter gave rise to several
definitions such as the phase, dwell, Larmor and traversal times among others. 
A short review of the evolution of these time concepts, 
followed by an account of experiments involving field-induced 
tunnel ionization and electron tunneling in a solid state junction is presented here.  
Whereas the former experiments use sophisticated techniques involving 
femtosecond laser pulses and determine the tunneling time by mapping the 
angle of rotation of the field vector to time, like the hands of a watch, 
the latter provides a simpler method through the measurement of current-voltage 
characteristics of the junction. 
\end{abstract}
\maketitle
\section{Introduction}
The definition of time has always
intrigued philosophers and physicists equally so.
Whereas in the opinion
of philosophers such as Immanuel Kant, space and time are the
framework within which the mind is constrained to construct its
experience of reality, a more pragmatic view was to consider time as
something that we use a clock to measure. In physics time appears as a
parameter, be it through Newton's second law, ${\bf F} = d{\bf p}/dt$,
in classical physics or
the Schr\"odinger equation, $i \hbar \partial \Psi/dt = H \Psi$,
in quantum mechanics. We may then ponder if
there is a way to measure time without referring to
the parametric time.
In other words, is there an expression which
represents a time interval without directly depending on
the parameter $t$? The answer to this question indeed leads us to
the quantum time concepts developed in connection with collisions
or scattering in three dimensions (3D) and tunneling in one dimension (1D).
As we shall see in the next sections, both these times in 3D and
1D are ``interaction times" of the subatomic particles involved. Their
definitions \cite{haugereview}
follow from similar conceptual considerations and find meaning
in physical processes \cite{weEPL,weAPL}.

\section{Evolution of quantum time concepts}
We are a long way from 1928 when Gamow published his pioneering
work \cite{gamow} on the tunneling of alpha particles in radioactive nuclei.
Though tunneling seems to be a well understood phenomenon 
with ramifications in many branches of physics, the amount of time 
spent by a particle in tunneling remains controversial. 
One of the earliest papers \cite{maccoll} on the topic studied the time evolution 
of a wave packet and concluded that there is no appreciable delay in 
the transmission of the packet through the barrier. 
Though the question of tunneling time as such did not attract much attention 
for another 25 years, it is interesting to note that the dwell time concept 
which was proposed by Smith \cite{smith} in 1960 appeared earlier in a different guise 
in a 1938 paper by Kapur and Peierls \cite{kapur} on the study of cross sections 
with resonances. 
The dwell time, $\tau_D$ (sometimes called residence time), is a stationary concept
and corresponds
to the time spent by a particle in a given region of space with interaction.
Smith derived the collision time in three dimensions (3D) and extended it to 
the multichannel case of elastic scattering with resonance formation. He constructed 
a lifetime matrix ${\bf Q}$ which was related to the scattering matrix ${\bf S}$ as, 
${\bf Q} = -i \hbar {\bf S} d{\bf S}^{\dagger}/dE$, such that the diagonal element 
$Q_{ii}$ gave the average lifetime of a collision beginning in the $i^{th}$ channel. 
In the one channel, elastic scattering case, this expression reduces to the phase 
time delay  
($\tilde{\tau}_{\phi}(E) = d\delta/dE$, with $\delta$ being the scattering phase shift) 
derived by Wigner and Eisenbud \cite{wigner} earlier. 
However, whereas the expression due to Smith which is 
derived from a time delayed radial wave packet is 
consistent with a lifetime matrix which is Hermitian, any Eisenbud-type 
lifetime matrix violates time reversal invariance \cite{wemultichannel}. 
The collision time 
of Smith reduces in one-dimension to the dwell time in tunneling \cite{buettikerPRB}
which is given as, 
$\tau_D(E) = \int_{x_1}^{x_2} \, |\Psi(x)|^2 \, dx/ j$, where, $|\Psi(x)|^2$ gives 
the probability density and $j$ the current density for a particle tunneling through a 
potential barrier with energy $E = (\hbar k)^2/2m$. 
It is then natural to expect a relation between the dwell time ($\tau_D$) and the phase 
time ($\tau_{\phi}$) which was 
indeed derived in \cite{winfulprl} and given by, 
\begin{equation}\label{phasedwelltime}
\tau_{\phi}(E) = \tau_D(E) - \hbar [ \Im m R/k]\, dk/dE\, . 
\end{equation}
It was shown in the 3D case to be \cite{meprl}, 
$\tilde{\tau}_{\phi}(E) = \tilde{\tau}_D(E) - \hbar \mu [ t_R/\pi]\, dk/dE$. 
The last term in (\ref{phasedwelltime}) arises due to the interference of the incident 
and the reflected waves in front of the barrier and makes the phase time singular 
near threshold. For large energies however, the phase and dwell times are the same. 
A relation between the phase time delay and number of resonances can be
found in \cite{srjain}.

In the years to follow, more definitions of tunneling time arose in different 
contexts. For example, considering the spin precession of an electron in a weak 
magnetic field, Buettiker \cite{buettikerPRB} defined the Larmor time which 
was related to the expectation value of the spin operator and reduced to the 
dwell time in the particular case of a rectangular barrier. 
The generalized Buettiker-Landauer time was derived later \cite{leavensaersSSC} 
and given by $\tau_{{\rm BL}} = -\hbar \, \partial ln|T|/\partial V$, 
which appears to be similar in form to the Pollak-Miller time \cite{pollak,alexandra1}, 
$\tau_{{\rm PM}} = \hbar \, \partial ln|T|/\partial E$. 
The four tunneling times: the Larmor time, Buettiker Landauer time, 
Wigner's phase time and Pollak-Miller time, 
originally derived from very different physical assumptions were 
derived in a unified manner within a Feynman path integral approach \cite{yamada} 
using Gell – Mann – Hartle decoherence functionals.
The total wave function was expressed as a
sum over all possible ``paths" with each path contributing a phase containing 
the action for that path.
At this point we refer the reader to the review articles \cite{haugereview,alexandra1, 
mugabooks,nussenzveig} 
and continue in the rest of the article to find out which of the 
tunneling time definitions correspond to physically measured times.

\section{Experimental extraction of electron tunneling times}
The time spent by subatomic particles in tunneling potential barriers
is usually estimated to be extremely small and beyond the reach of
experimental precision. Using the calculated values of free electron
Fermi energies and measured values of the work function,
Hartman \cite{hartman} estimated the phase times for metal - insulator - metal
sandwiches for several different materials to be of the order of
10$^{-16}$ s.

Though a direct measurement of such small times does not
seem feasible, the advent of intense laser fields has made measurements on
the tunneling of bound electrons from atoms possible \cite{eckle,nat1nat2,Pfeiffer}.
In \cite{eckle}, for example, an (intensity averaged) 
upper limit of 12 attoseconds on the
tunneling delay time in strong field ionization using helium atoms was placed.
More recently, the authors in \cite{Pfeiffer} find that the time delay in 
tunneling is zero for helium and argon atoms within the experimental 
uncertainty of a few 10s of attoseconds (10$^{-18}$ s).
In this strong field ionization process, the electron tunnels through the
potential created by a superposition of
the atomic Coulomb potential and the laser field. The
free electron is further accelerated by the laser field and
the tunneling time is determined by measuring the electron
momentum which depends on the strength of the field. 
In \cite{landsman} the authors perform a comparison of the extracted electron 
tunneling times with various theoretical definitions and conclude that 
only the Larmor time and the probability distribution of tunneling times 
constructed using a Feynman Path Integral formulation are compatible with experiment. 

Apart from these recent experiments, worth mentioning is also an earlier 
attempt \cite{fem} using
field emission microscopy (FEM). Measuring
the transversal momentum spread of the electrons
(with the help of a field emission microscope),
emitted by an isolated center of the tip,
the electron-tunneling time related with the field ionization of
this center could be deduced. The precision of the experiment lied
in the preparation of ultra sharp silicon tips (radius of curvature
10 - 20 nm) coated with 50 - 100 nm thick CaF$_2$:Sm$^{2+}$
layers which could be used as the field-emission
sources, for which the tunneling current would be due to the field
ionization of single isolated bivalent samarium dopant ions. 

Coming now to the tunneling of electrons through a solid state junction,
in \cite{weAPL} a novel method to extract the dwell times of electrons in
metal - insulator - metal sandwiches from current - voltage (I-V)
characteristics was presented. Tunneling of electrons in solid state junctions 
was studied earlier in \cite{gueretesteve}. 
Ref. \cite{weAPL} reported the I-V characteristics in
a Al/Al$_2$O$_3$/Al junction for temperatures ranging from 3.5 to 300 K.
The experimental data was then used to fit the barrier height and width
(for a rectangular barrier) using a standard semiclassical model 
for the I-V relation from \cite{simmons}.
The fits led to a constant value of barrier width
$s$$\sim$20.8~$\textrm{\AA}$ and a continuous increase in the barrier
height $V_0(T)$ from 1.799 eV at 300 K to 1.83 eV at 3.5 K.
Temperature dependence of the energy gap, $E_g(T)^{exp}$ was also determined
and allowed the authors to determine the average phonon frequency $\omega$.
An excellent fit to $E_g(T)^{exp}$ was obtained with an average phonon frequency,
$\omega$ = 2.05 $\times$ 10$^{13}$ sec$^{-1}$ \cite{weAPL}, 
in close agreement with the value
$\omega$ = 2.24 $\times$ 10$^{13}$ sec$^{-1}$, determined from the speed of sound
measurements using picosecond ultrasonic technique
in amorphous Al$_2$O$_3$. 
Having gained confidence about the precision of the measurements from the
above agreement of the phonon frequencies,
the barrier parameters were then used to
extract the temperature dependent dwell
times in tunneling. The average dwell time
$\tau_D$ was found to depend very weakly on temperature. The value of
$\tau_D$ was found to be 3.6 $\times$ 10$^{-16}$ sec
at mid-barrier energies. Extrapolating the values of the measured times in 
the field ionization experiments (see Fig. 3 of Ref. \cite{landsman}) to widths 
comparable to the above junction, the order of magnitude of the times is similar. 
The importance of \cite{weAPL} lies in the fact that knowledge of the 
tunneling time is obtained in a much simpler experiment as compared to \cite{landsman}. 
An extension of \cite{weAPL} including dissipative effects was done in \cite{weannals}. 

In passing, we note a recent experiment \cite{opticallat} based on the
merger of two fields \cite{morsch}
which have rapidly grown over the past few decades: optical
lattices (artificial crystals bound by light) and Bose Einstein Condensates (BEC).
In \cite{opticallat} a direct measurement of the tunneling delay time
through the barriers of an optical lattice was performed by studying the
time evolution of a Rubidium-87 BEC after a sudden displacement of the lattice.
The authors report delay times of the order of tens of $\mu s$.

\section{Concluding remarks}
We started the discussion in this article by asking if time
which appears as a parameter $t$ can also be expressed as a quantity which
does not explicitly depend on the parameter, $t$. Indeed, we saw different
quantum time concepts which are defined in terms of the energy,
wave function, energy dependent phases and flux of the tunneling particles.
Apart from the definitions themselves, there exist relations
which connect the times to quantities such as the density of states. The
energy derivative of the scattering phase shift which gives the
phase time delay for example, is given
via the Beth Uhlenbeck formula by \cite{ianocone},
\begin{equation}
\sum_l \, n_l(E) \, - \, n_l^0(E) \, =\, \sum_l {2 l + 1 \over \pi} \, 
{d\delta_l \over dE}\, , 
\end{equation}
where, $n_l(E)$ and $n_l^0(E)$ are the densities of states with and without
interaction respectively. This relation leads to interesting interpretations
involving ``time advancement" or negative time delay in a scattering process
involving unstable states \cite{metimeadv}. Similar relations can also be
derived in one dimension with the phase time defined in terms of the
phase of the transmission amplitude in tunneling \cite{gaspa}. 

Finally, we note that tunneling processes are inherently connected
to the survival and decay of unstable states \cite{francesco}.
This fact relates the tunneling times as well as the collision times to
the survival probabilities of the unstable states.
The half-life of a radioactive nucleus which mostly exhibits the
classical exponential
decay law can be shown to be given by the dwell time in the tunneling of
alpha particles through a Coulomb barrier \cite{weEPL}. The very same
dwell time concept can however be used to extract the non-exponential
behaviour of the decay law at large times \cite{wenonexpo}
as predicted by quantum mechanics \cite{urbanowski}.
Investigation of the quantum time concepts has indeed led to 
the understanding of time in its different guises.

\end{document}